\documentstyle
{article}

\author{Vadim R. Kudashev\\
Institute of Mathematics,\\
Chernyshevskii 112,\\
Ufa, 450000,\\
Russia\\
\\
E-mail: vadkud@nkc.bashkiria.su}
\title{KdV shock-like waves as invariant solutions \\
of KdV equation symmetries
}
\date{December 30, 1993
}

\begin{document}

\maketitle
\begin{abstract}
We consider the following hypothesis: some of KdV equation shock-like waves
are invariant with respect to the combination of the Galilean symmetry and
KdV equation higher symmetries. Also we demonstrate our approach on the
example of Burgers equation.
\end{abstract}

\section{Introduction}

\subsection{History of a problem [1-14]}

In [1] Gurevich and Pitaevskii (G-P) one of the first have formulated the
Korte\-weg-de Vries (KdV) equation shock-like waves problem: ''to find
solution of KdV equation

\begin{equation}
\begin{array}[b]{ll}
u_t+DRu\equiv u_t+uu_x+u_{xxx}=0, & u=u(x,t), \\
R=D^2+2u/3-D^{-1}u_x/3, & D\equiv d/dx,
\end{array}
\end{equation}
which has asymptotic behavior

\begin{equation}
\begin{array}{cc}
u^3=tu-x, & as\mid x\mid \rightarrow \infty ."
\end{array}
\end{equation}
In [2] Gurevich and Pitaevskii have supposed and investigated hypothesis:
the first term of asymptotic solution of (1), (2) is a continuous
combination of the ''external'' solution (2) and ''inner'' solution $%
u=\varphi (\theta )=\varphi (\theta +1):$%
\begin{equation}
\begin{array}{cc}
\varphi =2(r_3-r_1)dn^2(2K(m)\theta ;m)+r_1+r_2-r_3, & x^{-}(t)\leq x\leq
x^{-}(t),
\end{array}
\end{equation}
where $\theta _x=\kappa ,$ $\theta _t=\omega =-\kappa U$, $\kappa \cong
(r_3-r_1)^{1/2}/K(m)$, $U=(r_1+r_2+r_3)/3$, $m=(r_2-r_1)/(r_3-r_1),$ $dn$ -
is the Jacobi elliptic function, $K(m)$ - is the complete elliptic integral
of the first type, $r_1\leq 0,$ $r_1\leq r_2\leq r_3$, $r_3\geq 0$, and $%
r_{i}$ are governed by the Whitham-KdV (averaged) equations [3] (we
write these in form [4,5])

\begin{equation}
\begin{array}{lll}
r_{it}=\varphi ^i(r)r_{ix}, & r_i=r_i(t,x), & i=1,2,3, \\
\varphi ^i=(D_i\omega )/(D_i\kappa ), & D_i\equiv d/dr_i, &
\end{array}
\end{equation}
with appropriate boundary conditions. There are not rigorous theorems about
exact solution of (1), (2), but the non contradictoriness and
self-co-ordination of G-P hypothesis were proved in set works [2,5-11]. Note
also that closely connected questions were considered in [12].

The next step was made in [13,14] where it was shown that G-P solution is
simultaneous solution of non autonomous ordinary differential equation
(which is stationary part of KdV symmetry).

\subsection{The problem in question}

The problem in question is to find solutions of KdV equation (1) which have
asymptotic behavior

\begin{equation}
\begin{array}{cccc}
u^{2n+1}=tu-x, & as\mid x\mid \rightarrow \infty , & \mid t\mid \gg 1, &
n=1,2,...
\end{array}
\end{equation}
The analogous problem was considered in [4,7-11,15] with help of G-P-like
hypothesis for ''external'' and ''inner'' behavior of the first term for the
asymptotic solution.

Let us consider KdV equation symmetry (i.e. $u_{t\tau }=u_{\tau t},$ see,
for example, [16]) which is combination of KdV Galilean symmetry and KdV
higher order local symmetry

\begin{equation}
\begin{array}{cc}
u_\tau =(1-tu_x)+\gamma _mDR^mu, & m=2n, \\
\gamma _m=\prod\limits_{k=1}^m3(k+1)/(2k+1), & n=1,2,...
\end{array}
\end{equation}
The hypothesis in question is: solutions of (1), (5) are invariant with
respect to symmetry (6) i.e. $u_\tau =0,$ or

\begin{equation}
\begin{array}{ccc}
(1-tu_x)+\gamma _mDR^mu=0, & m=2n, & n=1,2,...
\end{array}
\end{equation}

The main goal of this Letter is to show the non contradictoriness and
self-co-ordination of this hypothesis with help of G-P-like form for the
first term of ''external'' and ''inner'' asymptotic solution of (1), (5),
(7). Also we well demonstrate our approach to problem on the example of
Burgers equation.

The importance of invariant solutions is well known. The main question is to
find such combination of symmetries which would be solve physically
interesting problem. It was ''linear, dispersible'' approach [17] which was
used in [13,14] to obtain and study equation (7) as $n=1$. In contrast with
[13,14] our suggestion is following: for studying the shock-like waves
problem it is useful to investigate such combination of symmetries which
would be solve problem in the ''dispersionless, nonlinear'' limit.

\section{External and single-phase inner solutions}

\subsection{''External'' solution}

Let us rewrite, for convenience, equations (1), (7) as

\begin{equation}
u_t+uu_x+\epsilon ^2u_{xxx}=0,
\end{equation}

\begin{equation}
\begin{array}{cc}
(1-tu_x)+\gamma _mD[\epsilon ^2D^2+2u/3-D^{-1}u_x/3]^mu=0, & m=2n,
\end{array}
\end{equation}
where $\epsilon $ is arbitrary constant. Our main observation: let $\mid
\epsilon \mid \ll 1$, ( the ''dispersionless, nonlinear'' limit) then for
any $m$ equations (8), (9) have formal external expansion

\begin{equation}
\begin{array}{ccc}
u=\sum\limits_{k=0}^\infty \epsilon ^{2k}u_k(x,t), & as\mid x\mid
\rightarrow \infty , & t\gg 1,
\end{array}
\end{equation}
with first term equals (5).

\subsection{''Inner'' solution}

In oscillation region we use solution in single-phase G-P-like form (3),
where solutions of (4) are defined by Tsarev's generalized hodograph method
[6] ($m=2n,$ $n=1,2,...)$

\begin{equation}
\begin{array}{cc}
\beta _{2m+3}\psi _{2m+3}^i(r)=\varphi ^i(r)t+x, & i=1,2,3,
\end{array}
\end{equation}
where $\beta _{2m+3}=3\gamma _m/(3+2m)$ and [4,5,10]
\begin{equation}
\begin{array}{cc}
\psi _{2m+3}^i=(D_i\omega _{2m+3})/(D_i\kappa ), & \omega _{2m+3}=-\kappa
U_{2m+3},
\end{array}
\end{equation}
with $U_{2m+1}$ (compare with [4]):
\begin{equation}
\begin{array}{cc}
LU_{2m+1}\equiv [\sum_k(2r_k^2D_k+r_k)]U_{2m+1}=3(m+1)U_{2m+3}, & U_1=1.
\end{array}
\end{equation}
Note useful expressions ($s=1,2,...$):
\begin{equation}
\begin{array}{l}
TU_{2s+1}\equiv (\sum_kD_k)U_{2s+1}=(2s+1)U_{2s-1}/3, \\
SU_{2s+1}\equiv (\sum_kr_kD_k)U_{2s+1}=sU_{2s+1}.
\end{array}
\end{equation}

As it is known Whitham-KdV equations (4) are averaged equations for (1),
(3). Substituting (3) into (7) and averaging over $\theta $ we obtain the
averaged equations for (3), (7) (compare with [4])
\begin{equation}
(1-tr_{ix})-\gamma _m\psi _{2m+1}^i(r)r_{ix}=0.
\end{equation}
Note that Whitham-KdV equations (4) admit symmetry
\begin{equation}
r_{i\tau }=(1-tr_{ix})-\gamma _m\psi _{2m+1}^i(r)r_{ix},
\end{equation}
which is combination of Whitham-KdV equations Galilean symmetry and
Whi\-tham-KdV equations higher order (nonclassical [16]) local symmetry. Thus
solutions of equations (4) which are described by (15) are invariant with
respect to symmetry (16).

Now we have questions: have equations (4), (15) the compatible solutions?,
what are these solutions?

{\it Proposition.} Equations (4), (15) are compatible, i.e. if we rewrite
(4) as
\begin{equation}
r_{it}=\varphi ^i(r)/(t+\gamma _m\psi _{2m+1}^i(r)),
\end{equation}
and (15) as
\begin{equation}
r_{ix}=1/(t+\gamma _m\psi _{2m+1}^i(r)),
\end{equation}
then $r_{itx}=r_{ixt}.$

{\it Remark.} This proposition is true for any semihamiltonian [6] systems, $%
\psi _i$ such that (compare with [6])
\begin{equation}
\begin{array}{cc}
D_k\psi ^i=(\psi ^k-\psi ^i)(D_k\varphi ^i)/(\varphi ^k-\varphi ^i), & i\neq
k,
\end{array}
\end{equation}
and $\varphi ^i$ such that $T\varphi ^i=-1.$

{\it Lemma.
\begin{equation}
\begin{array}{ccc}
T\psi _{2s+1}^i(r)=(2s+1)\psi _{2s-1}^i(r)/3, & s=1,2,...; & T\varphi ^i=-1.
\end{array}
\end{equation}
}

{\it Theorem.} Generalized hodograph solutions (11) satisfy equations (15).

{\it Proof. }The total derivative of (11) with respect to $x$ gives
\begin{equation}
\left. \sum\limits_kr_{kx}D_k[\beta _{2m+3}\psi _{2m+3}^i(r)-\varphi
^i(r)t]\right| _{(11)}=1.
\end{equation}
As it is known [6,9] the matrix in left side of (21) is diagonal
\begin{equation}
\left. D_k[\beta _{2m+3}\psi _{2m+3}^i(r)-\varphi ^i(r)t]\right|
_{(11)}=\delta _{ik}D_i\left. [\beta _{2m+3}\psi _{2m+3}^i(r)-\varphi
^i(r)t]\right| _{(11)},
\end{equation}
where $\delta _{ik}=1$ if $i=k$ and $\delta _{ik}=0$ if $i\neq k$. From
Lemma we have
\begin{equation}
T[\beta _{2m+3}\psi _{2m+3}^i(r)-\varphi ^i(r)t]=[\gamma _m\psi
_{2m+1}^i(r)+t].
\end{equation}
Combining (22) and (23) we obtain
\begin{equation}
\left. D_i[\beta _{2m+3}\psi _{2m+3}^i(r)-\varphi ^i(r)t]\right|
_{(11)}=\left. [\gamma _m\psi _{2m+1}^i(r)+t]\right| _{(11)}.
\end{equation}
Substituting (22), (24) in (21) we obtain%
$$
\left. r_{ix}[\gamma _m\psi _{2m+1}^i(r)+t]\right| _{(11)}=1.
$$
This is the same as (15). Thus generalized hodograph solutions (11) satisfy
(15).

{\it Remark.} The total derivative of (11) with respect to $t$ gives (17).
Note that the analogous theorem is formally true for any $\psi _\mu ^i$ from
(19) and such that $S\psi _\mu ^i=\mu \psi _\mu ^i$. It is followed from
expression $T\psi _\mu ^i=\gamma \psi _{\mu -1}^i$, which is followed from
commutation $(TS-ST)=T.$ The complete set of the appropriate $\psi _\mu ^i$
for any $\mu \geq -3/2$ $(\mu \neq -1/2)$ was obtained in [18].

\subsection{Statement}

{}From the above we obtain the main statement: the G-P-like first term (i.e.
continuous combination of ''external'' solution (5) and ''inner'' solution
(3), (11)) of asymptotic solution of (1), (5) satisfies (in framework of
G-P-like hypothesis) (7), (15) and may be considered as invariant solution
with respect to KdV equation symmetry (6). As it is followed from [8,11] if $%
m=2n$ then the plot of the multivalued function $\{r^{\mp },r_i\}$ is $C^1$
smooth near $x^{-}(t)$ and $x^{+}(t)$ (here $r^{-}$ is equal to $u$ from (5)
as $x\leq x^{-}(t)$, and $r^{+}$ is equal to $u$ from (5) as $x\geq
x^{+}(t)).$

\section{Exact solvable simplest examples}

\subsection{KdV equation case}

The simplest case of (7) is $(m=0,\gamma _0=1)$%
\begin{equation}
(1-tu_x)+u_x=0.
\end{equation}
{}From (25) we have solution of KdV equation%
$$
u=-x/(1-t),
$$
with ''initial profile'' $u(x,0)=-x$, and with singularity as $t=1.$

\subsection{Burgers equation case}

Now let us consider problem to find shock-wave solutions of Burgers equation
\begin{equation}
v_t+vv_x-v_{xx}=0,
\end{equation}
by the above symmetry approach. The first simplest case gives the same
solution as (25). The combination of Galilean symmetry and simplest higher
symmetry for (26) [16] gives us the following analog of (6)
\begin{equation}
v_\tau =D(x-tv)+D(4v_{xx}-6vv_x+v^3).
\end{equation}
Solution of (26) which is invariant with respect to symmetry (27) is defined
by equation
\begin{equation}
D(x-tv)+D(4v_{xx}-6vv_x+v^3)=0.
\end{equation}
Equation (28) has solution
\begin{equation}
(x-tv)+(4v_{xx}-6vv_x+v^3)=0.
\end{equation}
The ''external'' expansion for (29) has first term
\begin{equation}
v^3=tv-x.
\end{equation}
Substituting $v=-2V_x/V$ in (29) we obtain equation
\begin{equation}
8V_{xxx}-2tV_x-xV=0.
\end{equation}
{}From (31) we obtain the well known [19] solution of problem (26), (30) with
help of integral
\begin{equation}
V=\int\limits_{-\infty }^{+\infty }\exp (-\lambda ^4/8+t\lambda
^2/4-x\lambda /2)d\lambda .
\end{equation}

The combination of Galilean symmetry and fifth order symmetry of (26) [16]
gives equation
\begin{equation}
(x-tv)+(16v_{xxxx}-40vv_{xxx}-80v_xv_{xx}+40v^2v_{xx}+60vv_x^2-20v^3v_x+v^5)=0.
\end{equation}
Solution of (33) has external expansion with first term
\begin{equation}
v^5=tv-x.
\end{equation}
Substituting $v=-2V_x/V$ in (33) we obtain
\begin{equation}
32V_{xxxxx}-2tV_x-xV=0.
\end{equation}
The solution of (27), (33), (34), (35) is governed by integral%
$$
V=\int\limits_{-\infty }^{+\infty }\exp (-\lambda ^6/12+t\lambda
^2/4-x\lambda /2)d\lambda .
$$

{\bf Acknowlegment}

I am grateful to B.I.Suleimanov for many stimulating and useful discussions,
and to V.E.Adler, I.Yu.Cherdantsev, V.Yu.Novokshenov and V.V.Sokolov for
interest in this study. This work was supported, in part, by a Soros
Foundation Grant, and by RFFI grant 93-011-16088.

{\bf References}

[1] A.V.Gurevich and L.P.Pitaevskii, Zh. Eksp. Teor. Fiz. 60 (1971) 2155.

[2] A.V.Gurevich and L.P.Pitaevskii, JETP 38 (1974) 291.

[3] G.B.Whitham, Proc. Roy. Soc. A283 (1965) 238.

[4] V.R.Kudashev and S.E.Sharapov, The inheritance of KdV symmetries under
Whitham averaging and hydrodynamic symmetries of the Whitham equations,
preprint IAE-5221/6, Moscow

(1990) [in Russian]; Teor. Mat. Fiz. 87 (1991) 40.

[5] V.R.Kudashev, JETP Lett. 54 (1991) 175.

[6] S.P.Tsarev, Sov. Math. Dokl. 31 (1985) 488; Izv. Akad. Nauk SSSR, Ser.
Mat. 54 (1990)

1048.

[7] I.M.Krichever, Func. Anal. and its Appl. 22 (1988) 37.

[8] G.V.Potemin, Usp. Mat. Nauk 43 (1988) 211.

[9] B.A.Dubrovin and S.P.Novikov, Usp. Mat. Nauk 44 (1989) 29.

[10] F.R.Tian, Comm. Pure Appl. Math. 44 (1993) 1094.

[11] V.R.Kudashev, JETP Lett., 56 (1992) 320.

[12] P.D.Lax and C.D.Levermore, Comm. Pure Appl. Math. 36 (1983) 253, 571,
809; S.Venakides, Comm. Pure Appl. Math. 38 (1985) 883; R.F.Bikbaev and
V.Yu.Novokshenov, Proc. III International Workshop, Kiev, Naukova Dum\-ka,
1988, V.1, P.345.

[13] B.I.Suleimanov, Pis'ma Zh. Eksp. Teor. Fiz. 58 (1993).

[14] B.I.Suleimanov, Zh. Eksp. Teor. Fiz. (1994).

[15] A.V.Gurevich, A.L.Krylov and G.A.El', Pis'ma Zh. Eksp. Teor. Fiz. 54
(1991) 104.

[16] N.H.Ibragimov, Transformation Groups Applied to Mathematical Phy\-sics
(Reidel, Dordrecht, 1985); P.J.Olver, Application of Lie Groups to
Differential Equations (Springer, Berlin, 1986); A.C.Newell, Solitons in
mathematics and physics (Moscow, Mir, 1989).

[17] B.I.Suleimanov, L.T.Habibullin, Teor. Mat. Fiz. 97 (1993) 213.

[18] V.R.Kudashev, Phys. Lett. 171A (1992) 335.

[19] A.M.Ili'n, Matching of asymptotic expansions of solutions of boundary
value problems, ''Nauka'', Moscow, 1989; English transl., Amer. Math. Soc.,
Providence, RI, 1992.

\end{document}